# Ultrathin ultralight acoustic cloak

Igor I. Smolyaninov [1)] and Vera N. Smolyaninova [2)]

[1)] *Department of Electrical and Computer Engineering, University of Maryland, College Park, MD 20742, USA*

[2)] *Department of Physics Astronomy and Geosciences, Towson University, 8000 York Rd., Towson, MD 21252, USA*

**Existing designs of transformation acoustic cloaks are not easy to implement in many practical situations because of their large dimensions, while scattering cancellation cloaks do not protect the inner cloak volume. Here we report implementation of an acoustic metamaterial exhibiting "infinite cylinder-like" dispersion in an ultrathin (~1mm) and ultra-lightweight (~3g) acoustic cloak design intended to protect a human-head-size object in air, which combines scattering cancellation in the far-field with efficient sound proofing of the inner cloak volume.**

An acoustic cloak is a shell surrounding an object which lets a sound wave incident from any direction pass through and around the cloak without considerable absorption or scattering, thus making the cloak and the object acoustically "invisible". Acoustic cloaks share many common features with invisibility cloaks operating in various portions of electromagnetic spectrum [1,2]. Over the last few years there were quite a few theoretical and experimental demonstrations of acoustic cloaks, which typically operate based on either transformation acoustics [3-5], or scattering cancellation [6] designs. In addition to these "passive" designs, active cloaks, which use sound sources to cancel the incident wave are also being considered [7]. Unfortunately, transformation acoustic 3D and ground plane cloaks are difficult to implement in many situations, since limitations on the material parameters of the cloak lead to typical cloak dimensions being of the same order of magnitude as the cloaked region itself. Such an acoustic cloak would not be easy or practical to wear as a helmet if an application is sought in which a human head needs to be protected against an incoming intense acoustic wave front or a very loud sound. While this problem may be partially overcome with a scattering cancellation cloak, which typically requires a thinner shell around the cloaked object (compared to the transformation cloak), the scattering cancellation cloaks are designed to cancel the scattered wave in the near and far field outside the coated object, while they do not protect the inner cloak volume from the incoming wave. Here we report that these shortcomings of the existing acoustic cloaking technologies may be overcome using ultra-lightweight acoustic metamaterials exhibiting "'infinite cylinder-like" or "cylindrical" dispersion. These cylindrical acoustic metamaterials may be implemented in a design of an ultrathin and ultra-lightweight acoustic cloak, which combines scattering cancellation in the far-field with efficient sound proofing of the inner cloak volume. Potential applications of our design may include combat helmet and body armor, which may provide considerable protection of brain and lung tissues against harmful effects of intense sound and ultrasound waves and wave fronts.

Our design is based on an acoustic metamaterial, which has cylindrical dispersion, as illustrated in Fig.1a. The dispersion law of sound waves in an uniaxial anisotropic acoustic metamaterial may be written as

$$\omega^2 = k_z^2 c_z^2 + \left(k_x^2 + k_y^2\right)c_{xy}^2 \tag{1}$$

where $c_z$ is the velocity of sound along z direction, and $c_{xy}$ is the velocity of sound in the orthogonal directions. In the limit $c_z << c_{xy}$ the surface of equal frequency of phonons in the momentum space may be approximated as a cylindrical surface. Similar to layered (hyperbolic) electromagnetic metamaterials [8], this limit corresponds to a topological phase transition, so that acoustic metamaterial behaviour becomes quite unusual. Based on the Fermat principle, sound waves in such a material tend to propagate mostly in the xy plane. A practical implementation of such a metamaterial (Fig. 1b) may be based on thin layers of polyvinyl chloride (PVC, $c_{pvc}$= 1414 m/s) [9] separated by the layers of silica aerogel [10], which typically has much lower sound velocity of the order of $c_a$~ 50 to 100 m/s and extremely low density $\rho_a$~100 kg/m$^3$. We will demonstrate that such a metamaterial may be used to design an ultrathin (~1 mm) and ultralight (~3 g) wearable acoustic cloak for a human head.

As a first step, let us demonstrate that the ultra-low density cylindrical acoustic metamaterial shown schematically in Fig. 1 is indeed capable of excellent sound proofing. Silica aerogels are known to have excellent sound proofing properties on their own [11] due to considerable sound propagation losses. However, our theoretical simulations (Fig. 2) and test experiments clearly demonstrate that in accordance with the Fermat principle a layered cylindrical acoustic metamaterial structure indeed has considerably improved sound proofing properties compared to the bulk aerogels. The measured transmission of the fabricated 2 mm thick metamaterial sample which consists of 6 layers of 10 μm thick PVC film separated by layers of silica aerogel appears to be $9.2 \cdot 10^{-4}$ at 39.7kHz, which is ~12 times smaller than the transmission of the same thickness of the bulk aerogel.

Next, we will compare theoretical performance of 2D scattering cancellation cloaks designed using a conventional aerogel and a multilayer cylindrical PVC/aerogel metamaterial. These cloaks were designed to protect a 15 cm diameter cylinder made of material with density $\rho_w$=1000 kg/m$^3$ and the speed of sound $c_w$= 1500 m/s (a water-filled thin glass cylinder), which is placed in air. It is assumed that the average material parameters of a human head would also not deviate strongly from these values. The idea of our acoustic scattering cancellation cloak is similar to the operation principle of the scattering cancellation invisibility cloak [12], which is based on compensation of the positive polarizability $\sim(\varepsilon\text{-}1)V_d$ of a dielectric particle using a metal or metamaterial shell having negative polarizability ($\varepsilon<1$) and the appropriate volume $V_m$. Since the speed of sound in silica aerogel $c_a$ is smaller than the speed of sound in air ($c_0$=343m/s, $\rho_0$=1.225 kg/m$^3$), aerogel may be used to create a scattering cancellation cloak for a water cylinder in air, which would be somewhat similar to the operation of the electromagnetic cloak (since the speed of sound in water is larger than the speed of sound in air). However, we should point out that such an acoustic "negative polarizability" is useful but not necessary in the acoustic cloak design. Extensive work in the literature has shown that careful design and control of the density and bulk modulus (and shear modulus, in the case of elastic materials) all significantly affect the ability to achieve scattering cancellation, and may alleviate the need for "negative polarizability".

These alternative possibilities notwithstanding, the guiding principle of our design was an approximate mutual compensation of the "acoustic polarizabilities" of the component materials integrated over the volume of the cloak, followed by numerical fine tuning of the geometrical parameters of the cloak through scattering minimization. This numerical fine tuning was achieved by minimizing the scattering of a cloaked cylinder obtained by solving the 2D acoustic wave equation

$$\nabla \cdot \left( \frac{1}{\rho} \nabla p \right) - \frac{k_{eq}^2 p}{\rho} = 0 \quad , \tag{2}$$

where

$$k_{eq}^2 = \frac{\omega^2}{c^2} - k_z^2 , \tag{3}$$

and $p$ is the acoustic pressure, using the acoustic module of COMSOL Multiphysics 4.2a solver. Our simulation results are presented in Fig. 3. The scattering of a 6 kHz sound wave by an uncloaked 15 cm diameter cylinder shown in Fig. 3a produces considerable reflection and shadow. In addition, resonant enhancement of the sound wave can be observed inside the cylinder. Both reflection and shadow is considerably reduced in Fig. 3b, which was obtained by numerical optimization of the thin aerogel layer around the cylinder. The resonant enhancement inside the cylinder is also reduced. However, considerable "hot spots" of sound intensity may be still seen inside the cloaked cylinder. As we have mentioned above, such "hot spots" are bound to appear since the conventional scattering cancellation cloaks are designed to reduce the far-field scattering only. Finally, when the multilayer PVC/aerogel metamaterial is implemented (Fig. 3c), the far-field cloaking performance remains almost intact, while the sound intensity inside the cloak now stays near zero everywhere inside the cloaked cylinder. Thus, such a cloak combines scattering cancellation in the far-field with efficient sound proofing of the inner cloak volume, which is clearly evidenced by the comparison in Fig. 3d of the scattered field shown in the logarithmic scale from the cloaked and uncloaked cylinders. The left image in Fig.3d also illustrates the effect of material losses on the cloaking performance of our design. This image shows simulated performance of the cloak in which the aerogel losses are turned off numerically, with no apparent deterioration of performance.

We should also point out that the obtained cloaking behavior appears to be sufficiently broadband, which is apparent from theoretical simulations illustrated by Figs. 4 and 5. Fig. 4 illustrates our modeling of the acoustic cloak response to a typical shock wave. The original acoustic shock wave profile P(t) is taken from measurements by Dal Cengio Leonardi *et al.* reported in [13]. The acoustic cloak response to this shock wave is calculated as FFT$^{-1}$(C(ν)*FFT(P(t))), where P(t) is the pressure profile in a shock wave, FFT is the Fourier transformation, and C(ν) is the calculated frequency response of the acoustic cloak obtained from our COMSOL Multiphysics simulations. It appears that the cloak suppresses $\int Pdt$ in its interior by at least a factor of 100. These results are further verified in Fig.5, which shows the comparison of the simulated 7 kHz pulse propagation through an uncloaked and cloaked cylinder.

Experimental testing of scattering cancellation by the metamaterial acoustic cloak has been performed in the configuration shown in Fig. 6a. Comparison of the measured intensity of the scattered acoustic wave as a function of angle for the uncloaked and cloaked water-filled glass beakers is presented in Fig. 6b. In agreement with our numerical simulations, scattering suppression by approximately an order of magnitude has been indeed observed in these experiments.

The acoustic cloaking performance of the metamaterial-covered water-filled glass beaker was further tested using a commercial ultrasound motion sensor PASPort PS-2103A, produced by Pasco, as illustrated in Fig.7. The motion sensor transmits a burst of 16 ultrasonic pulses with a frequency of about 49 kHz. The ultrasonic pulses reflect off a target and return to the face of the sensor. After the transducer detects an echo, the sensor measures the time between the trigger rising edge and the echo rising edge. It uses this time and the speed of sound in air to calculate the distance to the object. As shown in Fig.7(a), the motion sensor is able to detect an uncloaked beaker located at a distance of 0.31 m in front of the sensor using ultrasonic pulses reflection off the beaker. On the other hand, when the beaker is cloaked and placed in the same location (see Fig.7(b)), the sensor cannot detect the presence of the cloaked beaker. It measures 3.16 m distance to a distant wall instead. This test experiment demonstrates an example of one possible application of the proposed acoustic cloaking scheme.

In conclusion, we have presented an implementation of acoustic metamaterials exhibiting "cylindrical" dispersion in an ultra-thin (~1mm) and ultra-lightweight (~3g) acoustic cloak configuration, which combines scattering cancellation in the far-field with efficient sound proofing of the inner cloak volume. The cloaking behaviour is broadband, so that it appears to be highly suitable in combat helmet and body armor applications, in which considerable protection of brain and lung tissues against harmful effects of intense sound and ultrasound waves may be provided.

We should also note that our technique follows in the footsteps of numerous important contributions to the field of acoustic cloaking, such as previous experimental results on scattering cancellation in air [14,15], and on the effect of absorption on a multilayer acoustic cloak [16]. We should also mention previous work on thin acoustic cloaking using scattering cancellation and hybrid designs [17,18], as well as various implementations of highly anisotropic multilayered metamaterial structures [19]. However, our choice of aerogel in the implementation of the layered acoustic metamaterials is critical, and it has not been studied previously. This choice enables very important practical applications due to the fact that the resulting cloaking designs are easily wearable, since they are extremely light and thin.

**Figure Captions**

**Figure 1.** The surface of equal frequency in momentum space (a) and schematic geometry (b) of ultra-low density "cylindrical" acoustic metamaterials.

**Figure 2.** (a) Comparison of sound wave propagation through bulk aerogel (left) and the multilayer PVC/aerogel metamaterial (right) of the same thickness. The sample boundaries are marked by arrows. The numerical simulations showing spatial distribution of acoustic pressure and its cross section were performed using COMSOL Multiphysics solver. (b) Photo of a fabricated metamaterial sample, which consists of 6 layers of PVC separated by layers of silica aerogel.

**Figure 3**. (a) Scattering of a 6 kHz sound wave by an uncloaked 15 cm diameter water-filled cylinder calculated using COMSOL Multiphysics 4.2a. Resonant enhancement of the sound wave can be observed inside the cylinder, as demonstrated in the inset. (b) Both reflection and shadow is considerably reduced by numerical optimization of the thin aerogel layer around the cylinder. However, considerable "hot spots" of sound intensity may be still seen inside the cloaked cylinder. (c) When the multilayer PVC/aerogel metamaterial is implemented, the far-field cloaking performance remains intact, while the sound intensity inside the cloak now stays near zero everywhere inside the cloaked cylinder. (d) Comparison of the scattered field shown in the same logarithmic scale from the cloaked and uncloaked cylinders. The left image shows simulated performance of the cloak in which the aerogel losses are turned off.

**Figure 4.** Modeling of acoustic cloak response to a typical shock wave: The original acoustic shock wave profile P(t) is taken from measurements reported in [13]. The acoustic cloak response to this shock wave is calculated as $FFT^{-1}(C(v)*FFT(P(t)))$, where P(t) is the pressure profile in a shock wave, FFT is the Fourier transformation, and C(v) is the calculated frequency response of the acoustic cloak.

**Figure 5**. Demonstration of the broadband behavior of the designed PVC/aerogel metamaterial acoustic cloak: comparison of the 7 kHz pulse propagation through an uncloaked (top row) and cloaked (bottom row) cylinders.

**Figure 6**. Experimental testing of scattering cancellation by the metamaterial acoustic cloak: (a) Experimental geometry. (b) Measured intensity of the scattered acoustic wave as a function of angle for the uncloaked and cloaked water-filled glass beaker.

**Figure 7**. Experimental testing of acoustic cloaking performance of the metamaterial-covered water-filled glass beaker using a commercial ultrasound motion sensor (PASPort PS-2103A, produced by Pasco): (a) The motion sensor detects an uncloaked beaker located at a distance of 0.31 m in front of the sensor using ultrasonic pulses reflection off the beaker. (b) When the beaker is cloaked and placed in the same location, the sensor cannot detect the presence of the cloaked beaker. It measures 3.16 m distance to a distant wall instead. (c) Zoomed in photo of the sample.

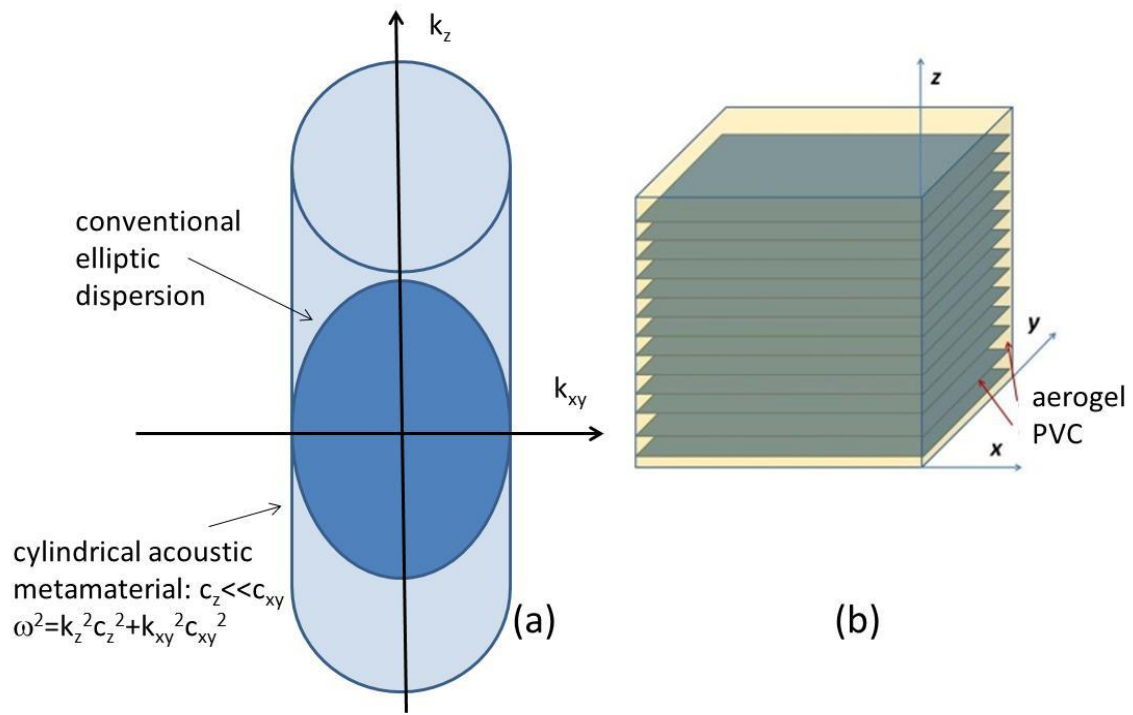

conventional elliptic dispersion

cylindrical acoustic metamaterial: $c_z \ll c_{xy}$
$\omega^2 = k_z^2 c_z^2 + k_{xy}^2 c_{xy}^2$

(a)

(b)

aerogel
PVC

Fig. 1

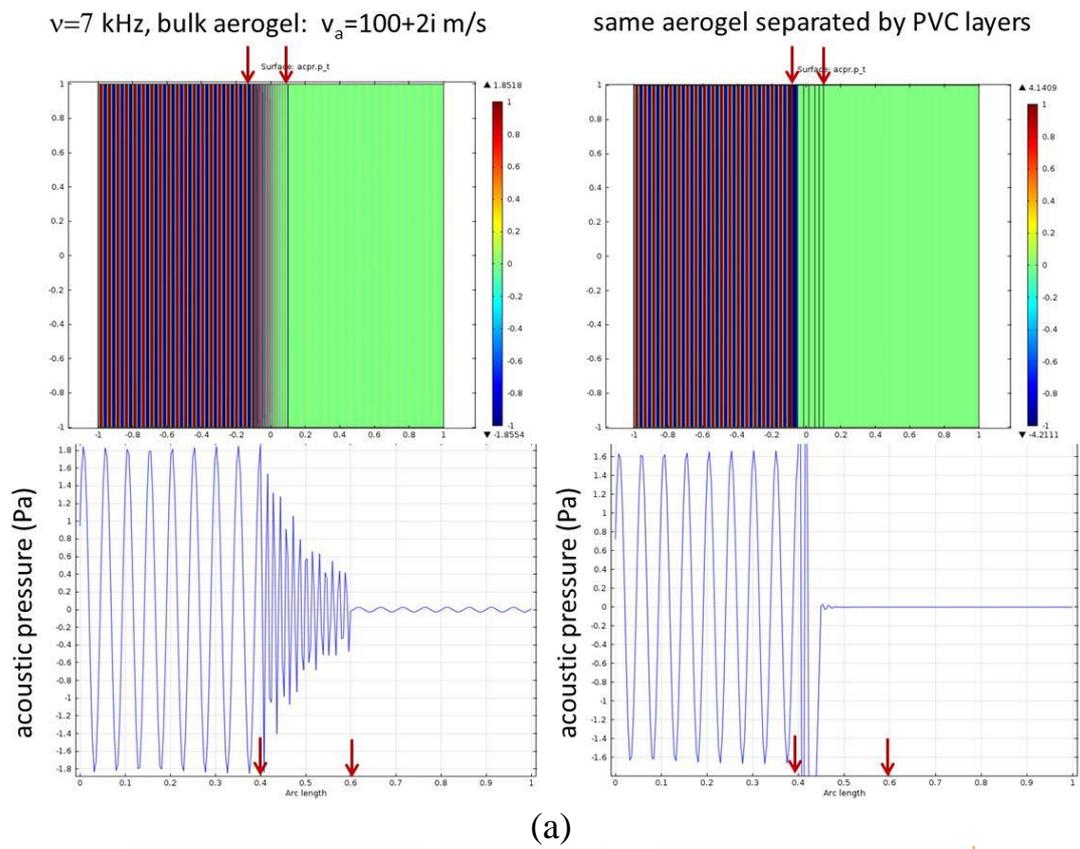

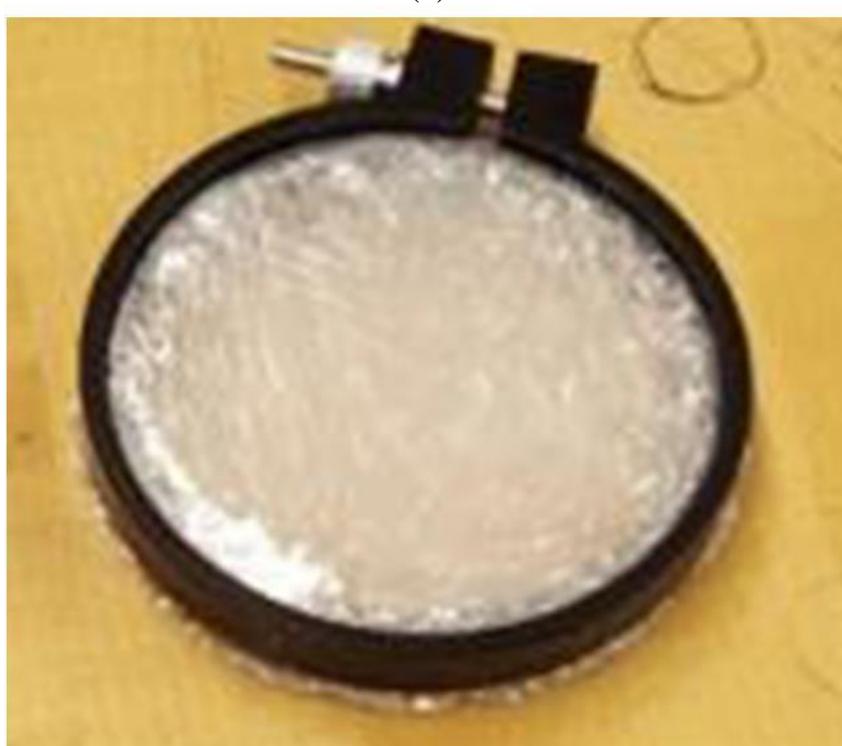

(a)

(b)

Fig. 2

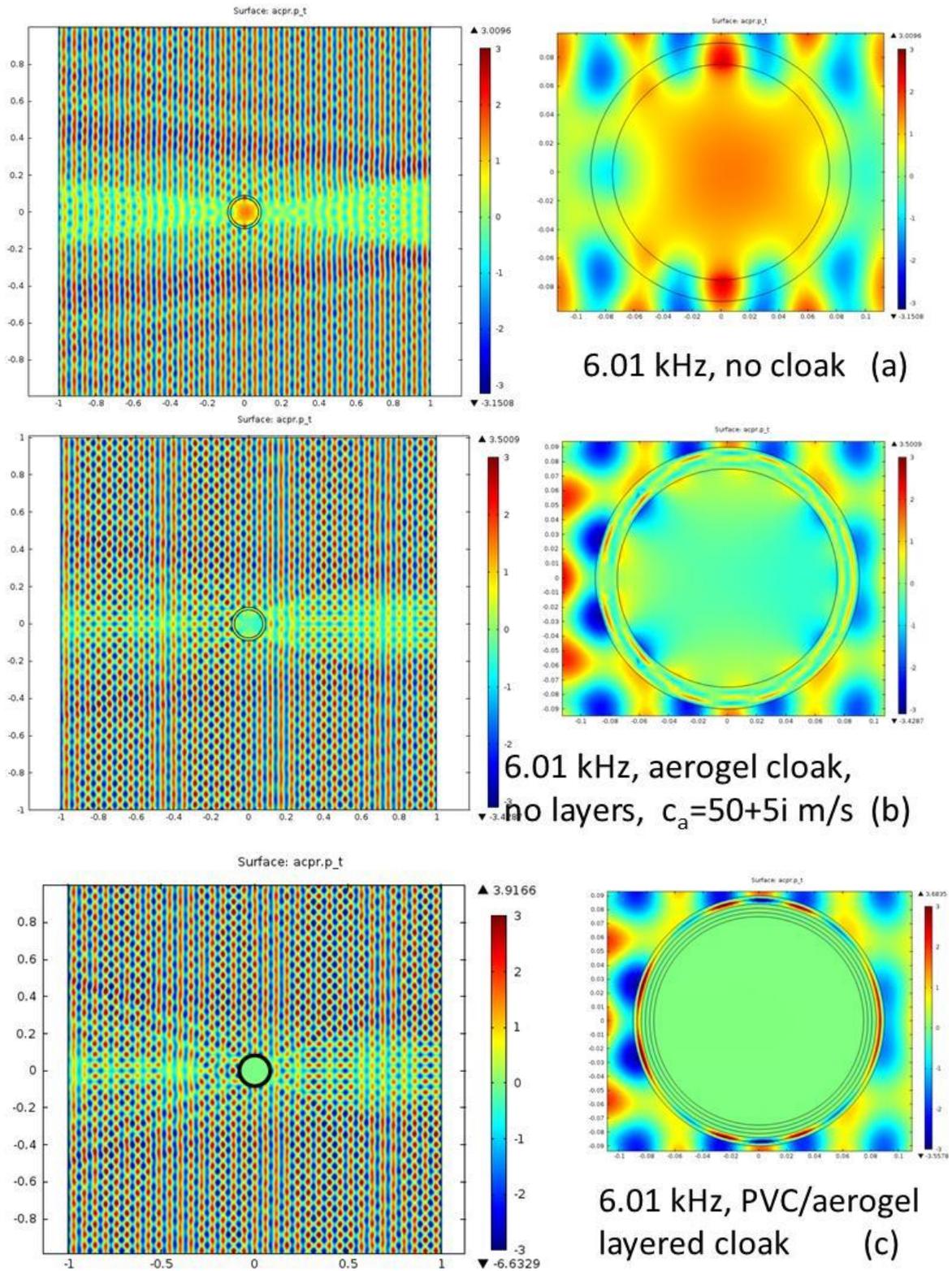

6.01 kHz, no cloak (a)

6.01 kHz, aerogel cloak, no layers, $c_a$=50+5i m/s (b)

6.01 kHz, PVC/aerogel layered cloak (c)

Fig. 3

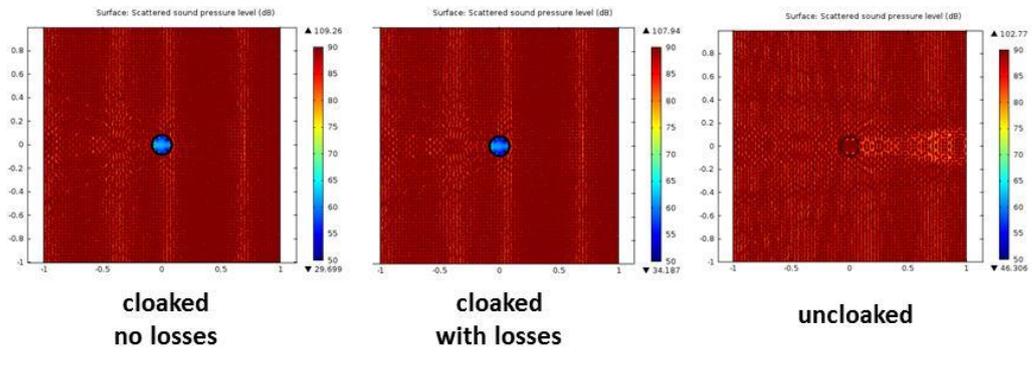

**cloaked no losses**  **cloaked with losses**  **uncloaked**

(d)

Fig. 3

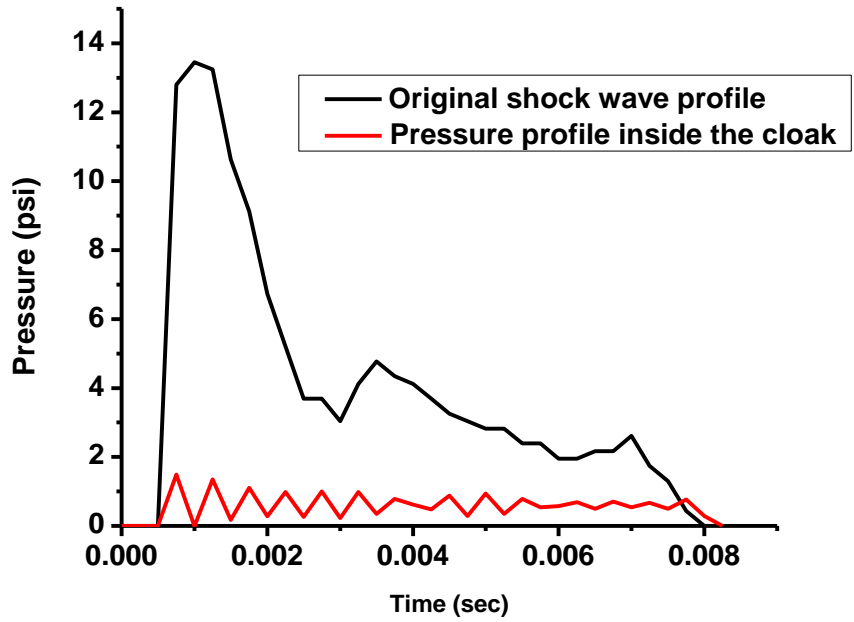

Fig. 4

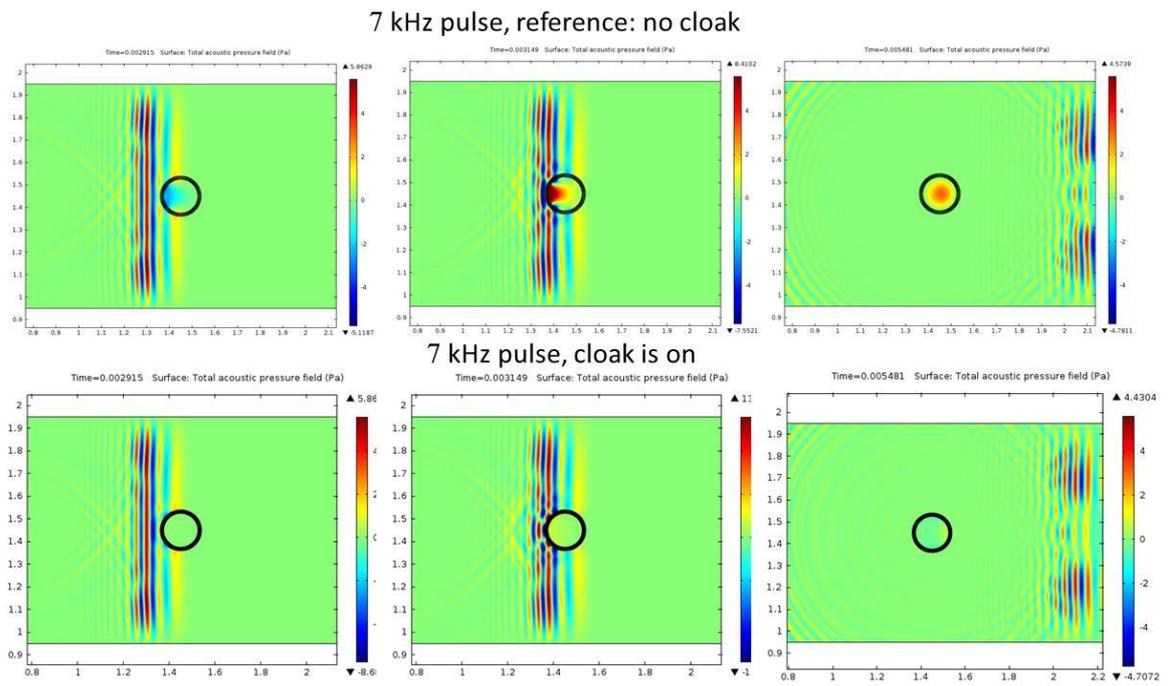

Fig. 5

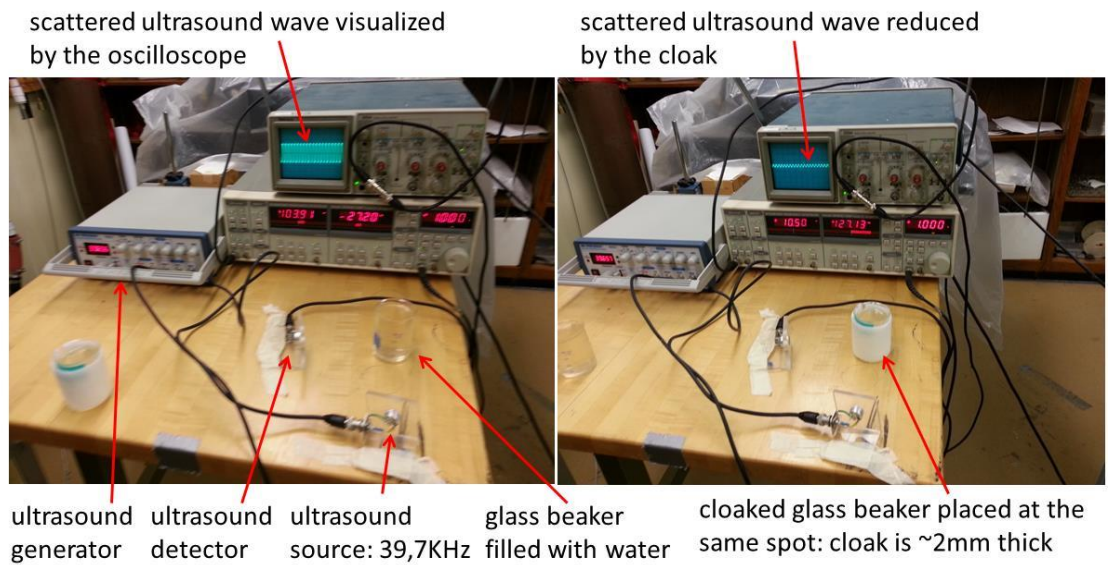

(a)

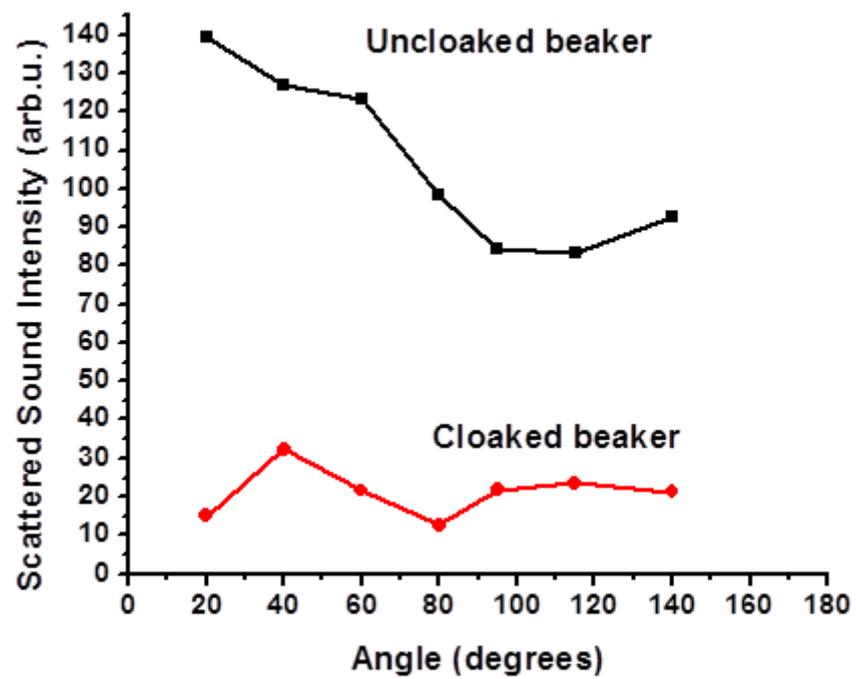

(b)

Fig. 6

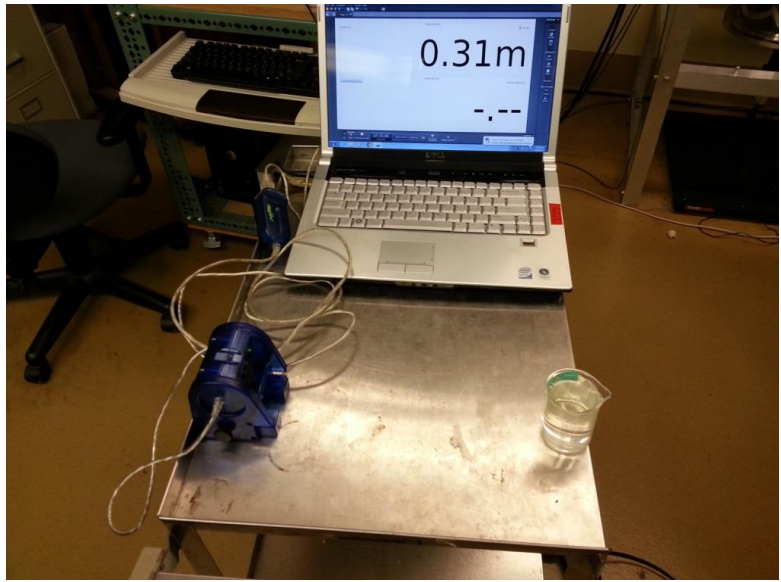

(a)

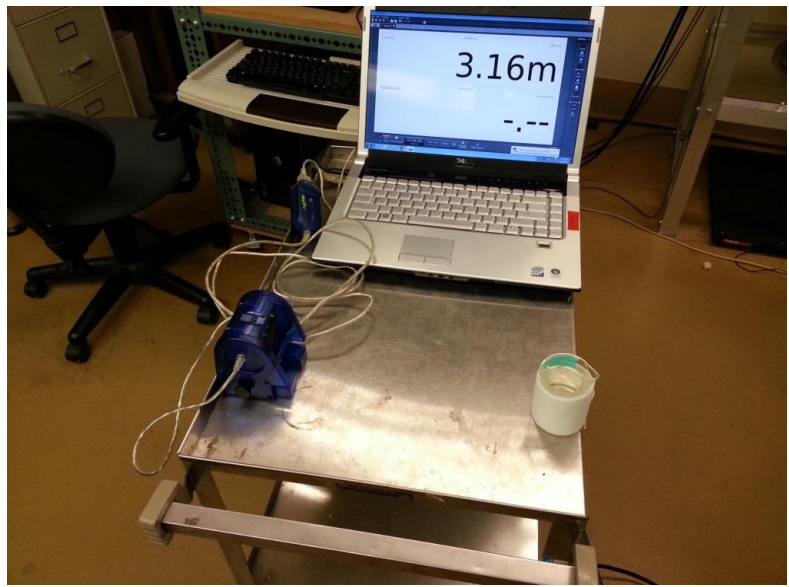

(b)

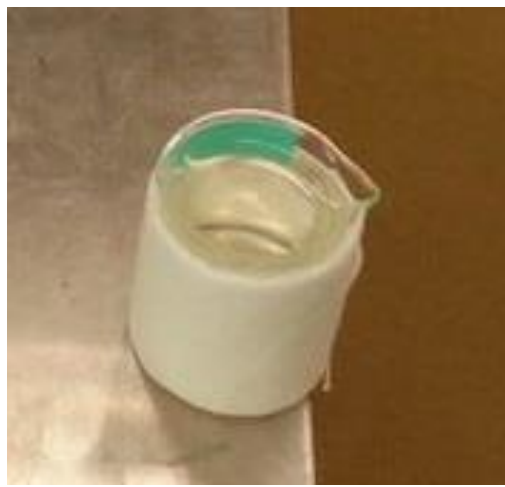

(c)

Fig. 7